# Vortex phase dynamics in yttrium superhydride YH$_6$ at megabar pressures

A.V. Sadakov[1,*], V.A. Vlasenko[1], I.A. Troyan[2], O.A. Sobolevskiy[1], D.V. Semenok[3,*], Di Zhou[3,*], V.M. Pudalov[1,4]

[1] P. N. Lebedev Physical Institute, Russian Academy of Sciences, Moscow 119991, Russia

[2] Shubnikov Institute of Crystallography, Federal Scientific Research Center Crystallography and Photonics, Russian Academy of Sciences, 59 Leninsky Prospekt, Moscow 119333, Russia

[3] Center for High Pressure Science & Technology Advanced Research, Bldg. #8E, ZPark, 10 Xibeiwang East Rd, Haidian District, Beijing, 100193, China

[4] National Research University Higher School of Economics, Moscow, 101000, Russia

*Corresponding authors: A.V. Sadakov (andrey.sadakov@gmail.com), D.V. Semenok (dmitrii.semenok@hpstar.ac.cn), D. Zhou (di.zhou@hpstar.ac.cn)

**Abstract**

A comprehensive study of the vortex phases and vortex dynamics is presented for a recently discovered high-temperature superconductor YH$_6$ with $T_C$ (onset) of 215 K under pressure of 200 GPa. Thermal activation energy ($U_0$) is derived in the framework of thermally activated flux flow (TAFF) theory. The activation energy yields a power law dependence $U_0 \propto H^\alpha$ on magnetic field with a possible crossover at a field around 8-10 T. Furthermore, we have depicted the vortex phase transition from vortex-glass to vortex-liquid state according to the vortex-glass theory. Finally, vortex phase diagram is constructed for the first time for superhydrides. Very high estimated values of flux flow barriers $U_0(H) = 1.5–7 \times 10^4$ K together with high crossover fields makes YH$_6$ a rather outstanding superconductor as compared to most cuprates and iron-based systems. The Ginzburg number for YH$_6$ $Gi = 3-7 \times 10^{-3}$ indicates that thermal fluctuations are not so strong and cannot broaden superconducting transitions in weak magnetic fields.

**Keywords:** Superconductivity; Hydrides; Vortex glass; Vortex liquid; TAFF; High pressure

**Highlights**

We present an in-depth study of the vortex matter dynamics in a high temperature superconductor (HTSC) $Im\bar{3}m$-YH$_6$ with transition temperature $T_C$ of 215 K under pressure of about 200 GPa. We establish the vortex glass and vortex liquid regions and analyze the dynamics of vortices within the thermally activated flux flow theory.

**Introduction**

Discovery of the superconductivity in H$_3$S system in 2015 [1] and subsequent works on superconducting LaH$_{10}$ [2,3], ThH$_{10}$ [4], YH$_6$ and YH$_9$ [5,6], (La,Y)H$_{10}$ [7], (La,Nd)H$_{10}$ [8] and several other hydride systems without a doubt heralded a new era in physics. A new family of high-temperature superconductors (HTSC) – high pressure hydrides – showed extreme superconducting properties: record breaking critical temperatures ($T_C$) up to 250 K [2,3], upper critical fields at zero temperature way above 100 Tesla [9] and currents carrying capabilities on par or exceeding the ones of commercial conventional superconducting wires [5].

In the past 8 years (2015-2023), since the discovery of hydride HTSC, some main characteristics of the superconducting state were firmly determined. It is commonly accepted that hydrides are phonon meditated type II superconductors, with small coherence length, large upper critical fields, and, seemingly, an s-wave order parameter. However, due to experimental limitations of diamond anvil cells quite a few fundamental superconducting properties are accessible and their peculiarities are not thoroughly investigated or even completely unexplored. One of the crucial topics, that still lacks experimental studies in the superconducting hydrides is the vortex state and the phase transitions within the vortex matter. Motivated by this gap of knowledge, we investigated the nature of vortex matter in YH$_6$ HTSC by comprehensive measurements and experimental data analysis.

It is well known that for a type II superconductor in the presence of an external magnetic fields, higher than the lower critical field, the ability to carry non-dissipative current is governed by the properties of vortices in the mixed state, originally named the Shubnikov state [10]. The dissipation starts when vortices begin moving and the counter to this motion is static disorder, which pins the vortex in a certain position and prevents it from movement. Generally, at low temperatures the vortices in the presence of random disorder form a vortex glass state [11,12]. This is a true superconducting state, capable of carrying non-dissipative current. When temperature increases and thermal vibration of the vortex position exceeds the significant part of the effective distance between vortices the system undergoes a phase transition into a liquid state, where the dynamics of the vortex liquid is regulated by thermal fluctuations. The magnitude of these fluctuations can be quantified by the Ginzburg number:

$$Gi = 10^{-9} \frac{\kappa^4 T_C^2}{H_{C2}(0)}, \qquad (1)$$

where $\kappa = \lambda/\xi$ is the Ginzburg-Landau parameter, $H_{C2}(0)$ is zero temperature upper critical field (in Oe), $\lambda$ and $\xi$ are London penetration depth and coherence length, respectively. In conventional type II superconductors, Ginzburg number is very small. For example, in Nb$_3$Sn $Gi = 1\times10^{-7}$ [13] in SnMo$_6$S$_8$, $Gi = 2\times10^{-6}$ [14].

In contrast, cuprate superconductors are characterized by very strong thermal fluctuations, and very high values of the Ginzburg number, $Gi = 1\times10^{-2}$ [12], which leads to a rich variety of vortex phases and vortex dynamics, the most evident of which is the significant broadening of the superconducting transitions in magnetic fields. In its turn, iron based HTSC on the other hand show quite a diversity in this regard. Some families of iron based superconductors (IBSC), like the "1111" family, have $Gi$ number as high as $1\times10^{-2}$, and also exhibit diverse vortex phases and vortex dynamics [15], and considerable broadening of the superconducting transitions [16,17]. Other families, like "11" or "122", on the contrary, have lower $Gi$ values (down to $3\times10^{-3}$ and $1\times10^{-4}$ respectively) [18], and much less transition broadening with field [16]. As for hydride HTSC, to our knowledge there are no estimations of the Ginzburg number, and very few studies on vortex matter in general, which justifies the relevance of our work as a significant contribution to the physics of high-Tc superconductivity.

In this work we present (i) a careful study of the charge transport properties for $Im\bar{3}m$-YH$_6$ superconductor in fields up to 16 Tesla and (ii) the data analysis within the framework of theories of thermally assisted flux flow (TAFF) and vortex glass (VG). As a result, we determined the activation energy of the vortex motion and its field dependence. Furthermore, we established the vortex glass region, where the vortices form a solidified glassy system, vortex liquid region, and the melting line between these domains.

**Results and discussion**

*1. Experimental details*

The superconducting YH$_6$ sample with $Im\bar{3}m$ structure was synthesized in a diamond anvil cell (DAC) by pulsed laser heating of yttrium particle in the ammonia borane media under pressure of ~200 GPa. Diameter of the diamond anvil was 50 μm. The details of the synthesis, structural XRD studies and the characterization of the sample were published elsewhere [5].

Measurements of electrical resistance and current-voltage characteristics were carried out with Keithley 2182a-6221 system with a standard four-probe technique. Cernox temperature sensor was fixed directly on a DAC very close to the sample with a thermoconductive paste in order to correctly measure temperature of the sample and avoid temperature hysteresis. Experiments in magnetic fields were performed with a Cryogenic CFMS-16 system.

*2. Thermally activated vortex motion*

As mentioned above, the region, where vortices start moving via thermally assisted activation mechanism is determined by the magnitude of Ginzburg number. Thus, studying the dynamics of the liquid state in superconducting hydrides we initially determine the value of Gi, defined by eq. (1). We take zero resistivity critical temperature, $T_c^{\text{offset}}$ = 205 K (from Figure 1a), and determine $H_{C2}(0)$ = 94 T and $\xi_0$ = 1.8 nm by fitting temperature dependence of the upper critical field $H_{C2}(T)$ with WHH model [19]. Value of the London penetration depth $\lambda(0)$ = 211 nm is derived from temperature dependence of the self-field critical current $J_{c\_SF}(T)$, in accordance with the model, proposed by Talantsev et al. [20,21]. The details of the $H_{C2}(T)$ and $J_{c\_SF}(T)$ fits are presented in the Supplementary Information, Figures S1 and S2. The obtained value of the London penetration depth is in a good agreement with recent theoretical estimations of $\lambda(0)$ = 150 nm for YH$_6$ system under pressure of 200 GPa [22]. As a result, we found the Ginzburg number $Gi$ = 3-7×10$^{-3}$, which is much higher than that for the majority of conventional superconductors, but still lower than those in cuprate HTSC and "1111" class of iron-based superconductors. This indicates that thermal fluctuations are not so strong in YH$_6$ and they are unlikely to cause significant broadening of the superconducting transitions in relatively weak magnetic fields. Indeed, as one can see in Figure 1a, the resistivity transition shifts to lower temperatures with applied field, but does not show significant broadening at least in fields up to 16 Tesla. In our opinion this clarifies the issue, raised by Hirsh and Marsiglio [23], that most of the current magneto-resistive data in hydrides do not show significant broadening of the superconducting transitions. However, it is worth mentioning, that this is the case only for relatively low fields, and when external fields are strong enough, resistive transitions start to broaden in superhydrides [8,24].

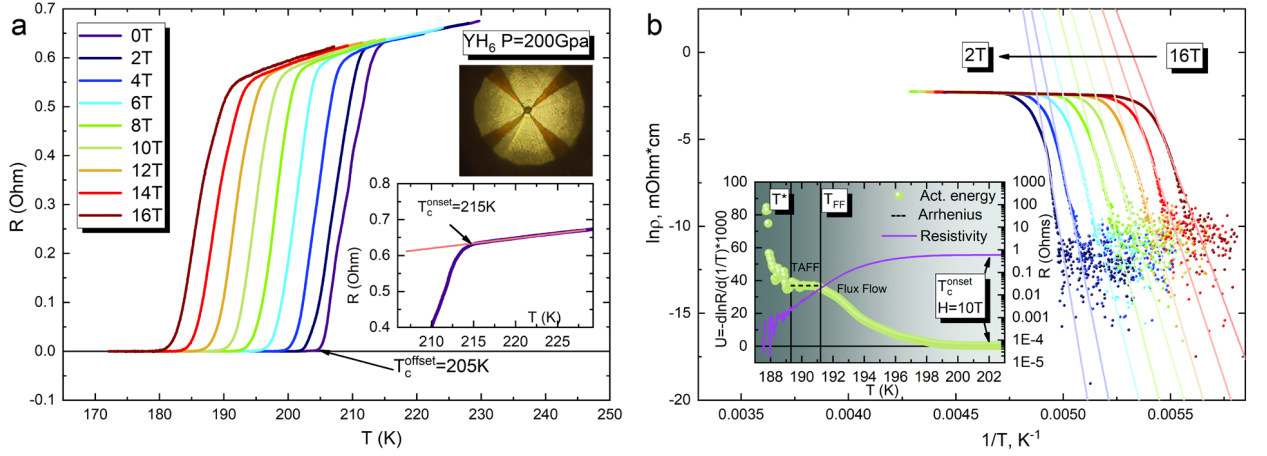

**Figure 1.** Resistive transitions in yttrium hydride YH$_6$ at high pressure. (a) Temperature dependence of the resistivity of YH$_6$ sample in magnetic fields of 0-16 T. The upper inset: chamber of the DAC with YH$_6$ sample and four electrodes. The lower inset: enlarged part of the transition onset. (b) Arrhenius plot ln(ρ) versus 1/T, where ρ – is resistivity of the sample with assumed thickness of 1.5 μm. The solid lines are fitting results from the Arrhenius relation. Inset: Temperature dependence of apparent activation energy u=-d(lnρ)/d(1/T) (left y-axis) and sample resistivity (right y-axis) for $H$ = 10 Tesla. T* depicts the lower temperature bound of the "flat" Arrhenius regime, T$^{FF}$ depicts the higher temperature bound of Arrhenius regime, above which the vortices are in the unpinned flux flow state.

In order to quantify the impact of thermal fluctuations in YH$_6$ we analyze our resistive data within the framework of TAFF theory [24]. In this regime, resistivity is expected to decrease exponentially as

$$\rho = \rho_0 e^{-U(T,H)/T}, \qquad (2)$$

where the activation energy signifies the energy required to de-pin a vortex [25]. Usually [15,26] $\rho_0$ is assumed to be temperature independent and the activation energy is assumed to be linear temperature dependent

$$U(H,T) = U_0(H) \times (1 - T/T_C), \qquad (3)$$

therefore, one can use a simplified relation

$$\ln(\rho) = \ln(\rho_0) + \frac{U_0(H)}{T_C} - \frac{U_0(H)}{T}, \qquad (4)$$

in order to determine U$_0$(H). In this case in the Arrhenius coordinates, ln(ρ) versus 1/T, a linear dependence corresponds to TAFF regime, and its slope gives us activation energy U$_0$(H). Arrhenius plot for YH$_6$ is presented in Figure 1b. The inset to Figure 1b shows temperature dependence of the apparent activation energy $U$ = – d(lnρ)/d(1/T) (left y-axis) and corresponding resistivity temperature dependence (right y-axis) for $H$ = 10 Tesla. Apparent activation energy reveals three regions. From onset transition temperature down to T$^{FF}$ we have the flux flow region. Here the activation energy remains relatively low and can be easily omitted, thus the vortex liquid is in the unpinned state. [27,28] The lower temperature bound of this regime is marked as T$^{FF}$ and with lowering temperature the derivative d(lnρ)/d(1/T) exhibits a plateau, corresponding to the Arrhenius regime and thermally activated flux flow. And the lower bound of the Arrhenius plateau is marked T*. At even lower temperatures, T < T*, d(ln ρ)/d(1/T) starts to grow abruptly, indicating that vortices in the liquid become strongly pinned [25,29]. Temperature dependence of apparent activation energy for various magnetic fields is presented in Supplementary Information, Figures S3. It is worth noting, that the Arrhenius region appears in all fields up to $H$ = 10 Tesla, and for higher fields temperature the derivative d(lnρ)/d(1/T) do not exhibit such a plateau. This is not uncommon for HTCS, for example in [30] authors showed that in 2212 cuprate apparent activation energy has an Arrhenius region in fields 1-5 Tesla and it vanishes in fields 12-22 Tesla. This qualitative change means, Arrhenius relation, based on the assumptions of linear temperature dependence of U(T,H) and temperature-independent $\rho_0$ may not be valid for fields $H > 10$ T.

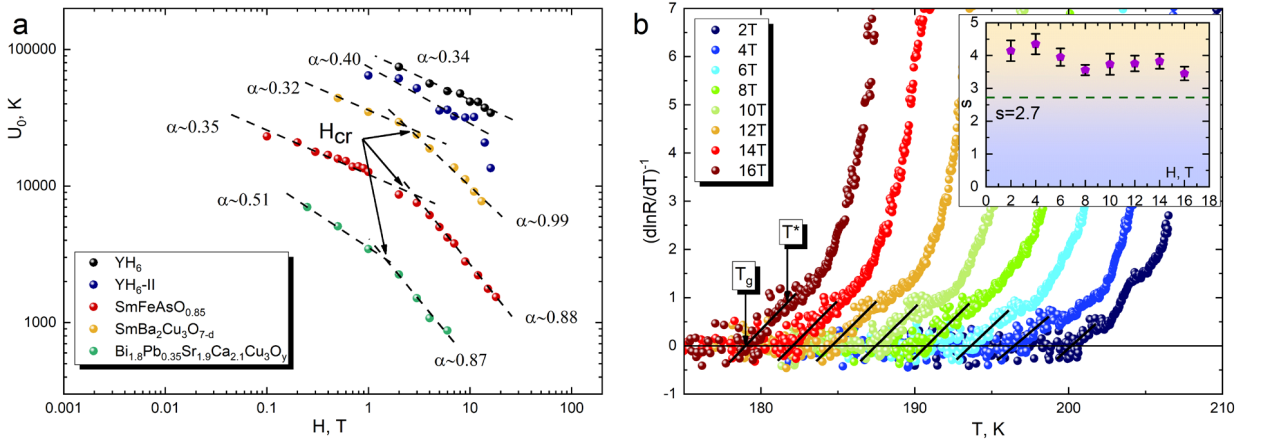

**Figure 2.** Activation energy and vortex glass-to-liquid phase transition in YH$_6$. (a) Field dependence of the activation energy $U_0(H)$ for two YH$_6$ samples in comparison with other superconductors. For comparison we present the data for iron based and copper-based superconductors SmFeAsO$_{0.85}$, SmBa$_2$Cu$_3$O$_{7-y}$ and Bi$_{1.8}$Pb$_{0.35}$Sr$_{1.9}$Ca$_{2.1}$Cu$_3$O$_x$. (b) Inverse logarithmic derivatives of resistivity data for different magnetic fields. Inset shows the obtained values of critical exponent $s$ (see eq. 5).

Figure 2a presents magnetic field dependence of $U_0(H)$, obtained from the linear parts of the Arrhenius plot (Figure 1b). Flux pinning energy exhibits a linear behavior in log-log coordinates, meaning that $U_0(H) \sim H^{-\alpha}$, with $\alpha = 0.34$. This value is very close to those for iron-based "1111" [15] and cuprate "123" superconductors [31] (shown for comparison in the Figure 2a) and is in a qualitative agreement with $U_0(H) \sim H^{-1/2}$, predicted by plastic flux creep theory [32], and related to the plastic deformation and entanglement of the vortices caused by point defects in the weakly pinned vortex-liquid phase. It is worth noting, that in cuprates and IBSC the $U_0(H)$ dependencies show a so called double-linear dependence, where in low fields power law dependence has $\alpha = 0.3 - 0.5$ and in higher fields $\alpha = 0.7 - 0.9$. The crossover field $H_{cr}$ is usually in the range of 2-4 T. For yttrium superhydride YH$_6$, linear behavior with $\alpha = 0.34$ continues to even stronger fields, and only shows a glimpse of a second linear slope at fields of ~10 Tesla. We checked the results for the second sample, YH$_6$-II (shown in the Figure 2a with dark blue symbols), which was less homogeneous. It turns out, that for YH$_6$-II the activation energy also seems to change the power law exponent from $\alpha = 0.40$ to higher value at $H > 10$ Tesla. This crossover is usually ascribed to a crossover from one pinning regime to another. Very high estimated values of flux flow barriers $U_0(H) = 1.5-7\times10^4$ K for both samples, together with very high crossover fields makes YH$_6$ a rather outstanding superconductor as compared to most cuprates and iron-based systems.

## 3. Vortex glass-vortex liquid phase transition and phase diagram

In order to get deeper insight into the nature of the vortex matter in YH$_6$ we performed a careful analysis of our data within the vortex glass theory. According to [33], resistivity of the type II superconductor in the vicinity of vortex liquid - vortex glass transition varies following a power law

$$\rho = \rho_0 \left|\frac{T}{T_g} - 1\right|^s, \qquad (5)$$

where $\rho_0$ is a characteristic resistivity, $s$ is the critical exponent related to vortex correlation length and its dynamics, $T_g$ is the vortex glass transition temperature. If we plot inverse logarithmic derivative of our low resistivity data versus temperature, power law becomes a straight line, which allows us to determine vortex glass temperature as an intercept with T-axis and critical exponent as a slope. This technique was successfully used to determine vortex glass transition lines in conventional superconductors [34], and all kinds of HTSC: iron-based superconductors [15,26], magnesium diboride [35], and cuprates [29].

Figure 2b shows a set of inverse logarithmic derivatives of the resistivity data for various magnetic fields. One can see a linear section, described by VG power law relation (2), from which we determine $T_g$ and $T^*$, a temperature, where the data deviates from linearity marking the upper temperature bound of this regime. The temperature $T^*$ is essentially the same temperature as one, determined in section 2. The inset in Figure 2b shows a field dependence of the obtained critical exponent ($s$) values. The values lie within range of 3.5-4.3, which, according to VG theory [33], corresponds to 3D vortex system, with $s = 2.7$ being a crossover value from 2D to 3D vortex system.

Furthermore, according to the modified vortex-glass model [36], the normalized resistivity can be rewritten as

$$\frac{\rho}{\rho_n} = \left[\frac{T(T_C-T_g)}{T_g(T_C-T)} - 1\right]^s, \qquad (6)$$

here $T_C$ values were taken as onset of resistive transition in magnetic fields, and $T_g$ values were taken from Figure 2b. Thus, according to Eq. (3), all the curves of normalized resistivity in different magnetic fields should scale to one curve in $\rho/\rho_0$ vs $T_{scale}$ coordinates ($T_{scale}$ is [$T(T_C - T_g)/T_g(T_C - T)$-1] and $\rho_0$ is the resistivity in normal state). Figure 3a shows that our data can be well scaled, confirming vortex glass region in YH$_6$. From the slope of the scaled data, we obtain a critical exponent, $s = 3.9 \pm 0.6$, which is in good agreement with the results, derived from VG theory [22]. The value of critical exponent lies above the value of 2.7, corresponding to 3D vortex model.

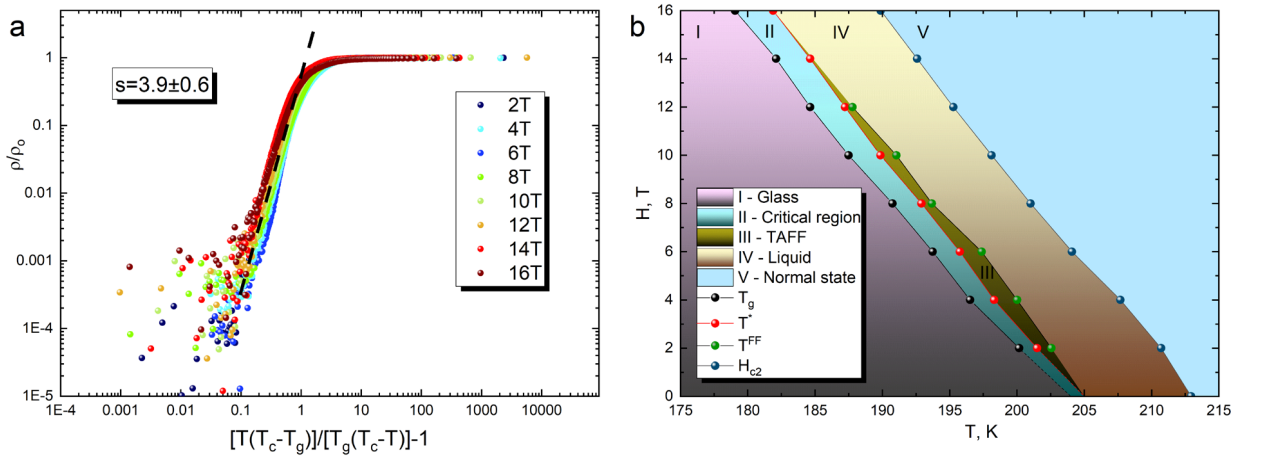

**Figure 3.** Fit of the experimental data based on the modified vortex-glass model, and the vortex phase diagram. (a) The normalized resistivities $\rho/\rho_0$ as a function of scaling temperature $T_{scale}$ in a double log plot. (b) Vortex phase diagram of YH$_6$ at about 200 GPa.

Finally, based on the values of $H_{c2}$ (T), $T_g$(H), $T^*$(H), $T^{FF}$(H) we plot the vortex phase diagram (Figure 3b), which exhibits five different regions: (I) vortex glass region, unfolding below the $T_g$(H) line; (II) vortex critical region, which stands between $T_g$(H) and $T^*$(H) lines, where the vortex matter gradually melts and a phase of pinned vortex liquid emerges; (III) TAFF region of the vortex liquid, spreading from $T^*$(H) line to $T^{FF}$(H) line; (IV) Flux flow region of the vortex liquid, between $T^{FF}$(H) and $H_{c2}$(T) lines; (V) normal state above $H_{c2}$(T) line.

**Conclusion**

To summarize, we have studied the vortex state in YH$_6$ polycrystalline sample under 2 Mbar pressure. Several vortex phases were established for the first time for compressed superhydrides. Within the framework of VG theory, we established the existence of vortex glass region, critical region and melting line $T_g$(H), separating these regions. The obtained values of critical exponent correspond to 3D vortex glass. Within the framework of TAFF theory we established that resistivity above critical region is thermally activated, with activation energy following U(T,H) ~ $H^α$×(1-T/T$_c$), where the power law parameter α = 0.34–0.40. This result is reminiscent of that in cuprates and iron-based superconductors. However, in YH$_6$ the magnitude of activations energies is considerably higher than in other HTSCs. Furthermore, weak power law decrease of U$_0$(H) continues in fields up to 10 Tesla, which goes beyond the usual for most HTSCs crossover field (2–4 Tesla), usually associated with a transition in vortex pinning regime. Finally, we depicted the vortex phase diagram.

The Ginzburg number for YH$_6$ is $Gi$ = 3-7×10$^{-3}$, which is much higher than that for the majority of conventional superconductors, but still lower than those in cuprate HTSC and "1111" class of iron-based superconductors. This indicates that thermal fluctuations are not so strong in YH$_6$ and may explain why superconducting transitions in high-$T_C$ polyhydrides do not broaden in weak magnetic fields.

**Acknowledgements**

A.V.S. thanks the support of RSF grant 22-22-00570.

**Data availability**

The authors declare that the main data supporting the findings of this study are contained within the paper and its associated Supplementary Information. All other relevant data are available from the corresponding author upon reasonable request.

# SUPPLEMENTARY MATERIALS

## Vortex phase dynamics in yttrium superhydride YH$_6$ at megabar pressures

A.V. Sadakov[1,*], V.A. Vlasenko[1], I.A. Troyan[2], O.A. Sobolevskiy[1], D.V. Semenok[3,*], Di Zhou[3,*], V.M. Pudalov[1]

[1] P. N. Lebedev Physical Institute, Russian Academy of Sciences, Moscow 119991, Russia

[2] Shubnikov Institute of Crystallography, Federal Scientific Research Center Crystallography and Photonics, Russian Academy of Sciences, 59 Leninsky Prospekt, Moscow 119333, Russia

[3] Center for High Pressure Science & Technology Advanced Research, Bldg. #8E, ZPark, 10 Xibeiwang East Rd, Haidian District, Beijing, 100193, China

[4] National Research University Higher School of Economics, Moscow, 101000, Russia

Corresponding authors: A.V. Sadakov (andrey.sadakov@gmail.com), D.V. Semenok (dmitrii.semenok@hpstar.ac.cn), D. Zhou (di.zhou@hpstar.ac.cn)

## Content

## 1. Coherence length determination

The estimation for zero temperature coherence length was made according to formula $H_{C2}(0)=\phi_0/2\pi\xi^2(0)$. And the temperature dependence for upper critical field down to zero temperature was calculated according to WHH model [1], see Figure S1.

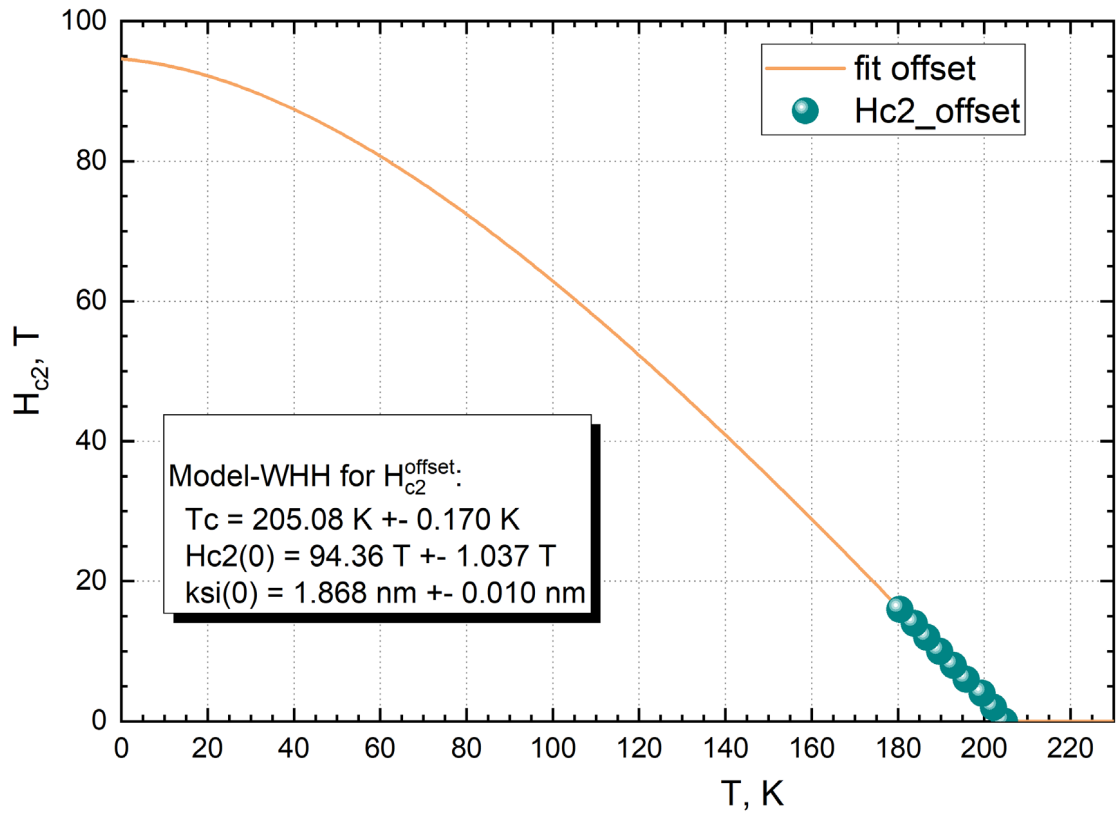

**Figure S1.** Werthamer–Helfand–Hohenberg fit for the upper critical field of YH$_6$ at ~200 GPa with the 'offset' criteria (zero resistivity point).

## 2. London penetration depth determination

As was originally proposed for thin films [2] and then developed for bulk samples [3] self-field critical current is a fundamental property of a superconductor and its temperature dependence carries the information about the London penetration depth and superconducting energy gap. According to [3] for type II superconductor of a rectangular cross-section shape self-field critical current is described as:

$$J_{c\_SF}(T) = \frac{\hbar}{4e\mu_0\lambda^3(T)}(\ln(\kappa) + 0.5) \cdot \left(\frac{\lambda(T)}{a}\tanh\left(\frac{a}{\lambda(T)}\right) + \frac{\lambda(T)}{b}\tanh\left(\frac{b}{\lambda(T)}\right)\right),$$

$$\frac{\lambda(T)}{\lambda(0)} = \sqrt{1 - \frac{1}{2k_BT}\int_0^\infty \cosh^{-2}\left(\frac{\sqrt{\varepsilon^2 + \Delta^2(T)}}{2k_BT}\right)}, \quad (S1)$$

$$\Delta(T) = \Delta(0) \cdot \tanh\left(\frac{\pi k_B T}{\Delta(0)}\sqrt{\eta\left(\frac{\Delta C}{C}\right)\left(\frac{T_C}{T} - 1\right)}\right),$$

where $2a$ – is the width of sample, $2b$ – is the thickness of sample, $\mu_0$ is the permeability of free space, e is the electron charge, $\kappa = \lambda/\xi$ is the Ginsburg-Landau parameter, $\xi$ is the superconducting coherence length, $\Delta(T)$ – the superconducting gap, $\eta = 2/3$ for s-wave superconductivity, and $\Delta C/C$ – is the specific heat capacity jump at the superconducting transition.

Finally, we get an approximation formula with only four fitting parameters: $T_C$, $\Delta(0)$, $\lambda(0)$, $\Delta C/C$. Our best fit with $T_C$ fixed at 206 K is presented in Figure S2 (yellow curve for $J_{c\_SF}(T)$, light blue for $\lambda(T)$). Derived penetration depth values are presented as well (dark blue symbols). The current-voltage experimental data is presented in the inset.

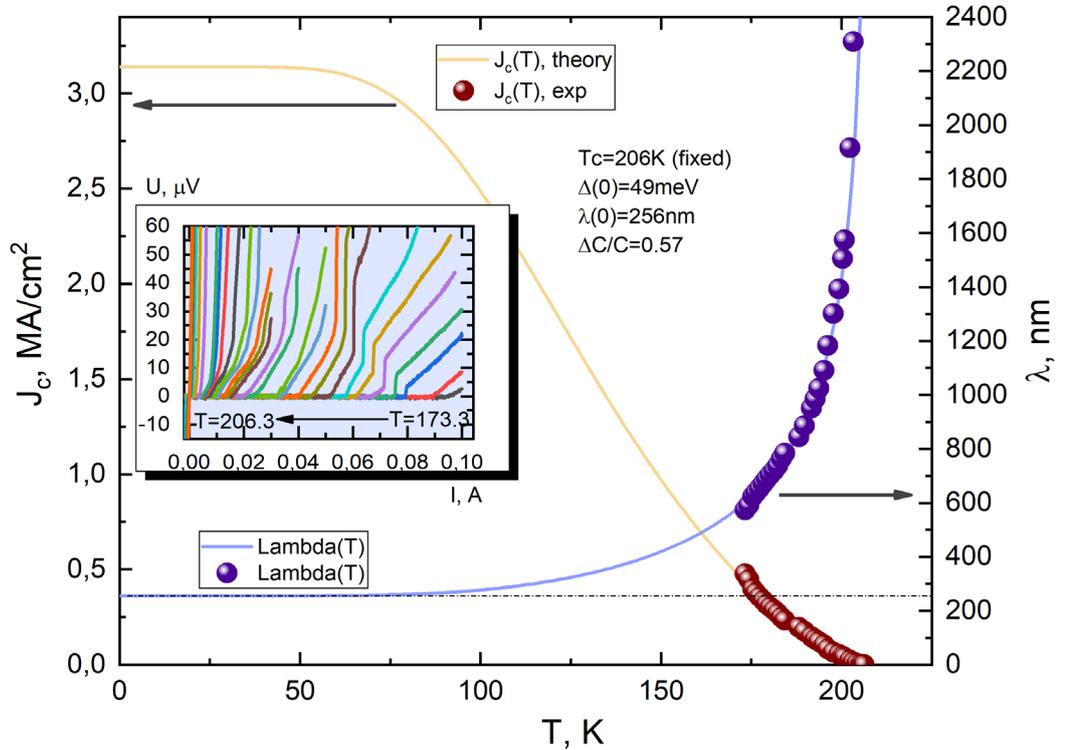

**Figure S2.** Temperature dependence of critical current density (red balls) and penetration depth (blue balls) and fitting curves based on the results of Talantsev et al. Inset: voltage-current (V-I) characteristic of YH$_6$ sample at about 200 GPa in the absence of magnetic field in the temperature range of 173 - 206 K.

### 3. Apparent activation energy in magnetic fields.

Figure S3 shows temperature dependence of the apparent activation energy $U = -d(\ln\rho)/d(1/T)$ for various magnetic fields. The dashed line depicts the position of the Arrhenius linear slopes. The vertical arrows point the the lower and upper bounds of the Arrhenius plateau. This flat region corresponds to thermally activated flux flow. As one can see, in fields $H > 12$ Tesla the Arrhenius region becomes very narrow, which, most probably, signifies the change in pinning regime. However, the nature of this change is yet to be studied.

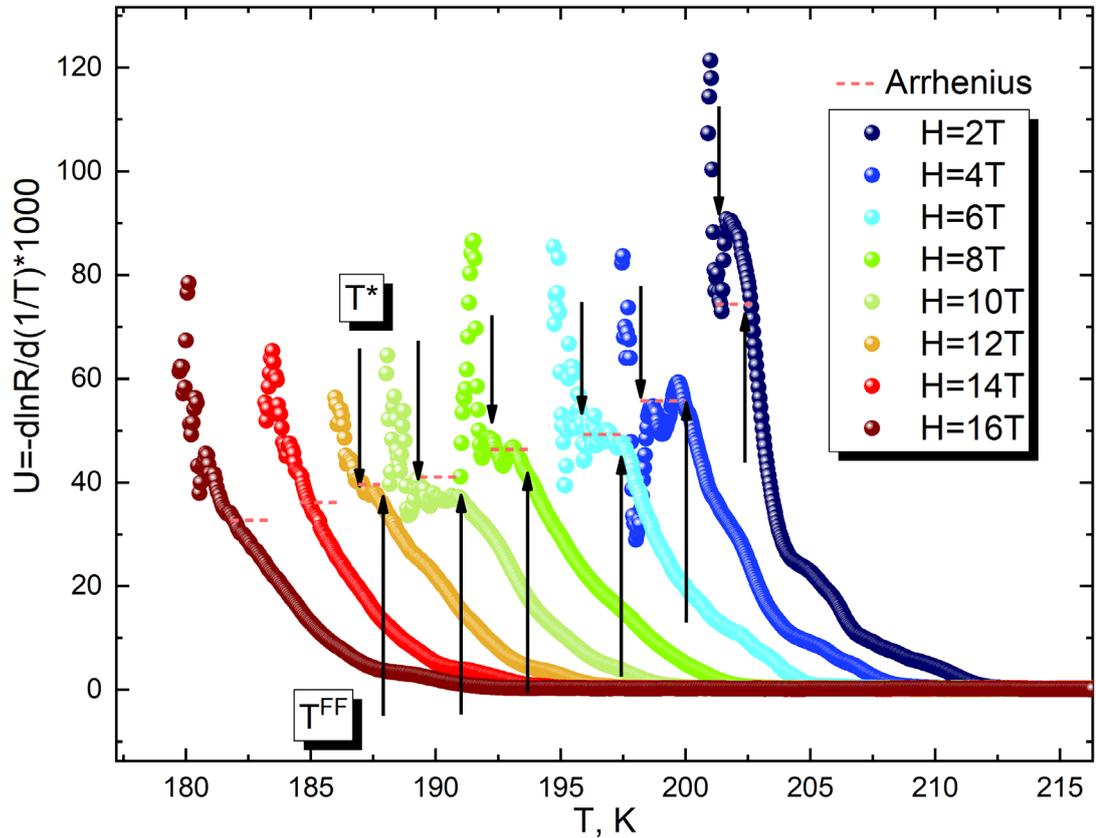

**Figure S3.** Temperature dependence of the apparent activation energy $U=-d(\ln\rho)/d(1/T)$ for various magnetic fields for $YH_6$. $T^*$ points to a lower temperature bound of the Arrhenius region, and $T^{FF}$ points to a higher temperature bound of this region. Dashed line shows the positions of the Arrhenius linear slopes.

## 4. Arrhenius plot for YH$_6$-II sample.

Second YH$_6$ sample is less homogeneous, and exhibits additional currently unexplained peculiarities in the lowest resistivity region nearby $T_C$, however it does still show a linear region in Arrhenius coordinates, and this region spreads for ~2 orders of magnitudes in resistivity.

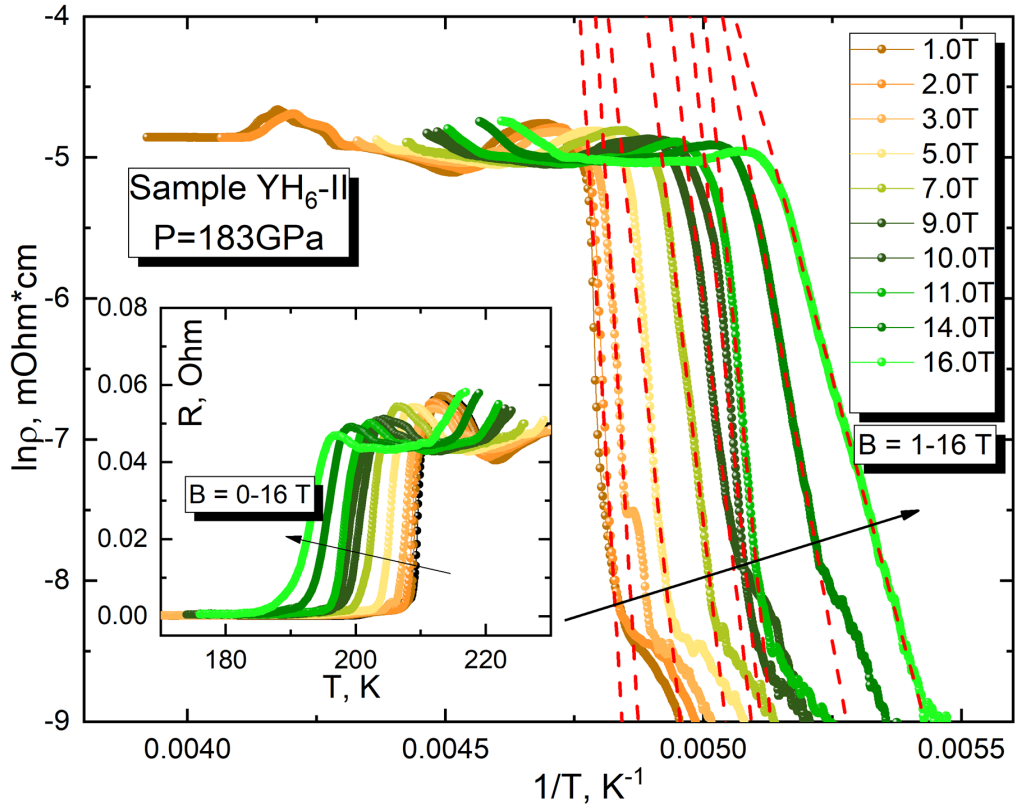

**Figure S4.** Arrhenius plot for YH$_6$-II sample in magnetic fields 1-16 Tesla. Dashed lines show the fits of the Arrhenius relation $\ln(\rho) \sim - U_0(H)/T$, where $U_0(H)$ is activation energy. Inset: superconducting transitions in the same series of magnetic fields.